\documentstyle[12pt,epsf]{article}
\begin{document}

\def\bb{b\bar{b}}
\def\bu{B^+}
\def\bd{B^0_d} 
\def\bs{B^0_s}
\def\lb{\Lambda_b}
\def\bmix{B^0 \mbox{--} \overline{B^0}}
\def\bdmix{B_d^0 \mbox{--} \overline{B_d^0}}
\def\bsmix{B_s^0 \mbox{--} \overline{B_s^0}}
\def\bsg{b\to s\,g}
\def\kstar{K^{\ast 0}}
\def\kstarbar{\overline{K}^{\ast 0}}
\def\sinsqth{\sin^2\theta_W^{eff}}
\def\Zbb{Z^0 \rightarrow b\,{\overline b}}
\def\Zcc{Z^0 \rightarrow c\,{\overline c}}
\def\Zff{Z^0 \rightarrow f\,{\overline f}}
\def\Zuds{Z^0 \rightarrow u\,{\overline u},d\,{\overline d},s\,{\overline s}}


\pagestyle{empty}

\renewcommand{\thefootnote}{\fnsymbol{footnote}}


\begin{flushright}
{\small
SLAC--PUB--7831\\
May 1998\\}
\end{flushright}

\begin{center}
{\large\bf B PHYSICS AT SLD: } \\[1mm]
{\large\bf $\bu$ and $\bd$ LIFETIMES, $\bs$ MIXING}\\[1mm]
{\large\bf AND SEARCH FOR $\bsg$ DECAYS}\footnote{Work supported by
Department of Energy contract  DE--AC03--76SF00515.}

\vspace{1.cm}

{\bf St\'ephane Willocq}

Stanford Linear Accelerator Center

Stanford University, Stanford, CA 94309

\vspace{3.mm}
{\it Representing}
\vspace{.2cm}

{\bf The SLD Collaboration}

\end{center}

\vspace{6.cm}   

\begin{abstract}
We report new preliminary $B$ Physics results obtained with 300,000 hadronic
$Z^0$ decays collected by the SLD experiment at the SLC between 1993 and 1997.
Three analyses are presented: a measurement of
$\bu$ and $\bd$ lifetimes, a study of the time dependence of
$\bsmix$ mixing, and a search for decays of the type $\bsg$.
All analyses benefit from the small and stable interaction point
and the excellent resolution and efficiency provided by the
pixel-based CCD Vertex Detector.
The $\bsg$ analysis also exploits the particle identification capabilities
of the Cherenkov Ring Imaging Detector.
\end{abstract}

\vfill
\vspace*{1cm}

\begin{center} 
{\it Invited talk presented at the
33rd Rencontres de Moriond:} \\
{\it Electroweak Interactions and Unified Theories} \\
{\it Les Arcs, France }\\
{\it 14-21 March 1998}\\
\end{center}

\vfill\eject

\baselineskip=18pt    

\noindent {\large \bf 1.  Introduction}
\vspace*{3mm}

In this paper, we focus primarily on $B$ decay studies
for which $e^+ e^-$ colliders
at the $Z^0$ pole are particularly well-suited due to the high $B$ hadron
boost: lifetimes and time-dependent mixing.
We also report the results of an inclusive search for $\bsg$ decays.
Furthermore, the SLD experiment is particularly well-equipped
to study $B$ decays.
First, the SLC interaction region is small and stable,
with transverse dimensions of the order of 1~$\mu$m.
Second, the small interaction point (IP) can be {\em tracked} using
the combined information from the Central Drift Chamber and the
pixel-based CCD Vertex Detector (VXD) with an
uncertainty of 4.4~$\mu$m transverse to the beam direction.
The IP position along the beam axis is
measured with an accuracy of 15 (35) $\mu$m for $\Zuds$ ($\Zbb$) decays.
The impact parameter resolution at high momentum is
determined from $Z^0 \rightarrow \mu^+ \mu^-$ decays to be
$\sigma(r\phi) = 11$ $\mu$m and $\sigma(rz) = 24$ $\mu$m.
Multiple scattering yields an additional momentum-dependent
contribution parameterized as
$\sigma = 33\, \mu\mbox{m} / (p~\sin^{3/2}\theta)$,
where the momentum $p$ is expressed in GeV/c.
Note that the above describes the performance of the new vertex
detector (VXD3) installed prior to the start of the 1996 run.
For the performance of the previous detector (VXD2), as well as a general
introduction to the SLD detector, see Ref.~\cite{Rb}.


All measurements presented below are preliminary.

%
%

\vspace*{3mm}
\noindent {\large \bf  2.  $\bu$ and $\bd$ Lifetimes}
\vspace*{3mm}

The study of exclusive $B$ hadron lifetimes provides an important test
of our understanding of $B$ hadron decay dynamics.
Lifetimes are especially useful to probe the
strong interaction effects arising from the fact that $b$ quarks
are not free particles but are confined inside hadrons.
In the naive spectator model, the $b$ quarks are treated
as if they were free and one therefore expects
$\tau(\bu) = \tau(\bd) = \tau(\bs) = \tau(\lb)$.
However, the measured charm hadron lifetimes follow the pattern
$\tau(D^+)\simeq 2.3~\tau(D_s)\simeq 2.5~\tau(D^0)\simeq 5~\tau(\Lambda_c^+)$.
These factors are predicted to scale with the inverse
of the heavy quark mass squared and
the $B$ hadron lifetimes are thus expected to differ by only
10-20\%~\cite{Bigi,Neubert}.

  The measurement technique used by SLD takes advantage of the excellent 3-D
vertexing capabilities of the VXD to reconstruct the decays
inclusively. The goal is to reconstruct and identify all the
charged particles originating from the $B$ decay chain. This then allows
charged and neutral $B$ mesons to be separated by simply measuring
the total charge of tracks associated with the $B$ decay.

  The analysis~\cite{blrtopol} uses an inclusive topological vertexing
technique~\cite{ZVTOP} to tag and reconstruct $B$ decays.
Secondary vertices are found in 65\% of $b$ hemispheres but in only
20\% of $c$ hemispheres and in less than 1\% of $uds$ hemispheres (for VXD3).
The $B$ sample purity is increased by reconstructing the vertex mass $M$,
which includes a partial correction for missing decay products.
Requiring $M > 2$ GeV yields a $B$ hadron sample with 98\% purity
and 50\% efficiency.
In the hadronic $Z^0$ event sample collected between 1993 and 1997,
we select 35,947 $B$ decay candidates
by requiring the vertices to have a mass $M > 2$ GeV,
a decay length $L > 1$ mm, and a transverse distance from the
IP $< 2.4$ (2.2) cm for VXD2 (VXD3).
\begin{figure}[t]
  \hspace*{10mm}
  \epsfxsize=14cm
  \epsfbox{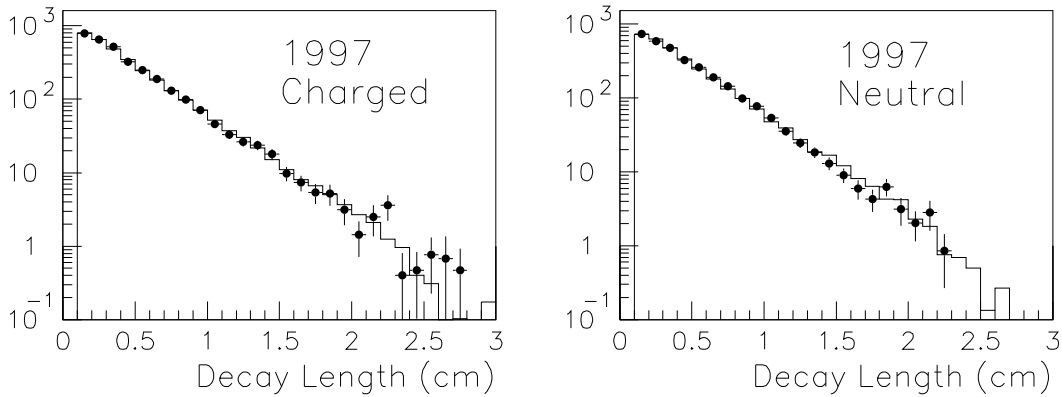}
  \caption{\it \label{fig_decl}
  \baselineskip=12pt
  Decay length distributions in the charged and neutral samples
  for 1997 data (points) and best-fit Monte Carlo (histograms).}
  \baselineskip=18pt
\end{figure}
The sample comprises 14,064 neutral and 21,883 charged vertices
corresponding to reconstructed decays with total charge
$Q = 0$ and $Q = \pm 1, 2, 3$,
respectively, where $Q$ is the charge sum of all tracks associated with
the vertex. Monte Carlo (MC) studies show that
the ratio between the number of $\bu$ and $\bd$ decays
in the charged sample is 1.55 (1.72)
for VXD2 (VXD3), and the ratio between
$\bd$ and $\bu$ decays in the neutral sample is 1.96 (2.24) for
VXD2 (VXD3)\footnote{Reference to a specific state
(e.g., $B^+$) implicitly includes its charge conjugate (i.e., $B^-$).}.

  The $\bu$ and $\bd$ lifetimes are extracted with a simultaneous
binned maximum likelihood fit to the decay length distributions of
the charged and neutral samples (see Fig.~\ref{fig_decl}).
The fit yields lifetimes of
$\tau_{\bu} = 1.665\pm0.029(\mbox{stat})\pm0.042(\mbox{syst})$ ps,
$\tau_{\bd} = 1.612\pm0.030(\mbox{stat})\pm0.055(\mbox{syst})$ ps,
with a lifetime ratio of
$\tau_{\bu}/\tau_{\bd}
    = 1.030^{+0.035}_{-0.033}(\mbox{stat})\pm0.027(\mbox{syst})$.
The main contributions to the systematic error on the ratio
come from uncertainties
in the detector modeling, $\bs$ lifetime, $b$-baryon fraction,
fit systematics, and MC statistics.
These measurements are the most statistically precise to date
and confirm the expectation that the $\bu$ and $\bd$ lifetimes
are nearly equal.

\vspace*{3mm}
\noindent {\large \bf  3.  $B^0_s - \overline{B^0_s}$ Mixing}
\vspace*{3mm}

  Transitions between flavor states $B^0$ and $\overline{B^0}$
take place via second order weak interactions ``box diagrams.''
A measurement of the oscillation frequency $\Delta m_d$ for $\bdmix$ mixing
determines, in principle, the value of the Cabibbo-Kobayashi-Maskawa (CKM)
matrix element $\left| V_{td} \right|$, which in turn gives information on the
Standard Model (SM) CP-violating phase $\eta$ and the parameter $\rho$---both
of which are currently poorly constrained.
However, theoretical uncertainties in calculating hadronic matrix elements
are large ($\sim 25\%$~\cite{Paganini})
and thus limit the current usefulness of precise $\Delta m_d$ measurements.
These uncertainties are significantly reduced ($\sim$ 6--10\%)
for the ratio between $\Delta m_d$ and $\Delta m_s$.
Thus, combining measurements of the oscillation frequency of
both $\bdmix$ and $\bsmix$ mixing translates into a measurement of
the ratio $|V_{td}| / |V_{ts}|$ and provides a strong constraint
on the CKM parameters $\rho$ and $\eta$.

  Experimentally, a measurement of the time dependence of $\bmix$
mixing requires three ingredients: (i) the $B$ decay proper time has
to be reconstructed, (ii) the $B$ flavor at production
(initial state $t = 0$) needs to be determined, as well as (iii) the $B$
flavor at decay (final state $t = t_{\rm{decay}}$).
At SLD, the time dependence of $\bsmix$ mixing has been studied using
three different methods described below: lepton+``D'', lepton+tracks,
and Charge Dipole.
All three use the same initial state flavor tag
but differ by the method used to either reconstruct the $B$ decay and/or
tag its final state flavor.
The data consists of some 150,000 hadronic $Z^0$ decays collected with
VXD3 in 1996 and 1997.

\begin{figure}[thb]
  \vspace*{4mm}
  \hspace*{30mm}
  \epsfxsize=9cm
  \epsfbox{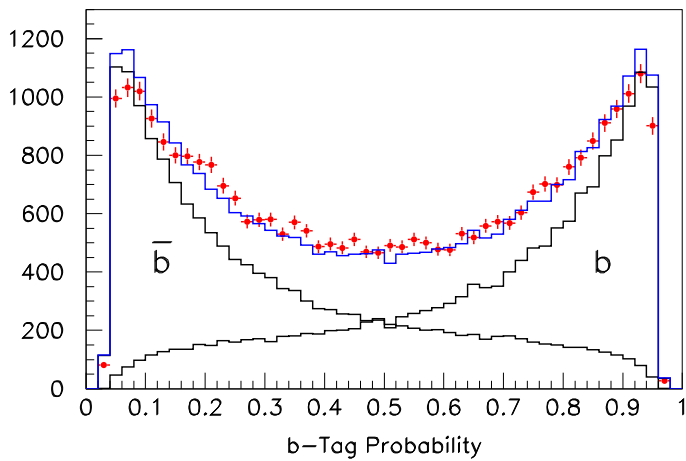}
  \caption{\it \label{fig_initag}
  \baselineskip=12pt
  Distribution of the computed initial state $b$-quark probability for
  1997 data (points) and Monte Carlo (histograms) showing the $b$
  and $\bar{b}$ components.}
  \baselineskip=18pt
\end{figure}

  Initial state tagging takes advantage of the large
longitudinal polarization of the electron beam ($\sim 74\%$) and the
pronounced polarization-dependent forward-backward asymmetry in $\Zbb$ decays.
For left- (right-) handed electrons and
forward (backward) $B$ decay vertices, the initial quark is tagged as
a $b$ quark; otherwise, it is tagged as a $\overline{b}$ quark.
The initial state tag is augmented by the following information from the
hemisphere opposite that of the reconstructed $B$ vertex:
momentum-weighted track charge, vertex charge, vertex charge dipole,
kaon charge and lepton charge.
These various tags are
combined to yield an initial state tag with 100\% efficiency and
effective average right-tag probability of 85\%.
Figure~\ref{fig_initag} shows the computed $b$-quark probability for data
and MC simulation, and clearly illustrates the separation between $b$ and
$\overline{b}$ components.

\begin{figure}[thb]
  \vspace*{-5mm}
  \hspace*{30mm}
  \epsfxsize9cm
  \epsfbox{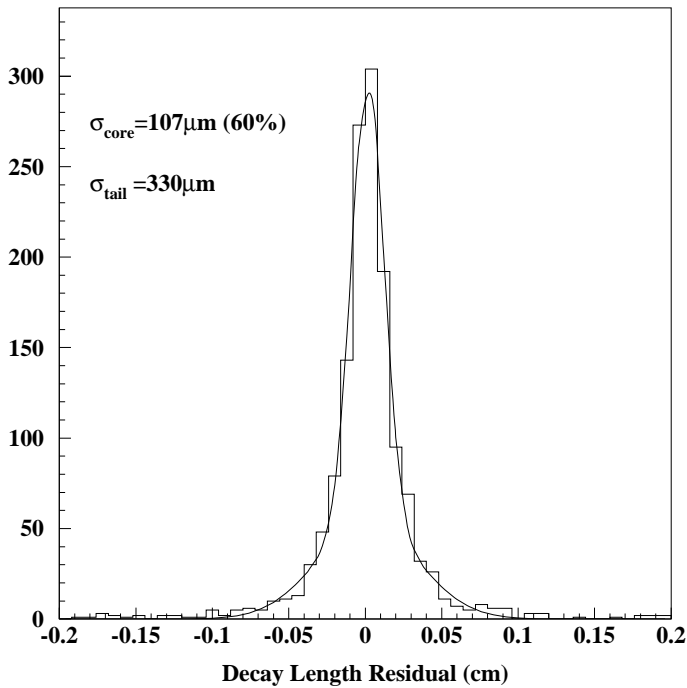}
  \caption{\it \label{fig_resol}
  \baselineskip=12pt
  Distribution of the decay length residual
  for the lepton+``D'' analysis. Also shown is a fit
  with a double Gaussian.}
  \baselineskip=18pt
\end{figure}

  The lepton+``D'' analysis aims at reconstructing the charged track $B$ and
$D$ vertex topologies of semileptonic $B$ decays.
It proceeds by first selecting event hemispheres
containing an identified lepton ($e$ or $\mu$).
Then, a ``D'' vertex is reconstructed using a similar topological
technique as that used for the lifetime analysis described earlier.
This vertex is constrained to lie near the plane containing the lepton
and the IP, and to be downstream of the lepton, thereby reducing the
confusion between primary and secondary tracks and thus allowing
efficient reconstruction of semileptonic $B$ decays at short decay lengths.
Several cuts are added to clean up the $D$ vertex candidate and
reduce the contamination from cascade charm semileptonic decays
($b \to c \to l$).
In particular, the lepton momentum transverse to the $D$ trajectory
is required to be greater than 0.9 GeV/c
and the $\chi^2$ for fitting the lepton and $D$ vertex tracks to a
single vertex is required to be larger than that obtained for the
$D$ vertex tracks alone.
The $B$ decay vertex is reconstructed by intersecting the lepton and $D$
trajectories.
The decay length resolution for direct $(b \to l$) decays
is shown in Fig.~\ref{fig_resol}.
Reconstruction of the $B$ hadron boost uses both tracking and calorimeter
information.
The proper time resolution is a function of the proper time $t$:
$\sigma_t = \left[(\frac{\sigma_L}{\gamma\beta c})^2
 + (t\,\frac{\sigma_{\gamma\beta}}{\gamma\beta})^2\right]^{1/2}$;
it is dominated by the decay length resolution $\sigma_L$
at small proper time with Gaussian widths of $\sigma_t \simeq 0.06$ ps
(60\% fraction) and 0.18 ps, whereas at large proper time it is dominated
by the boost resolution $\sigma_{\gamma\beta}$ and grows with increasing $t$.
In order to enhance the $\bs$ fraction in the sample, the total charge of
all tracks associated with the decay is required to be 0.
For this and the other analyses, the $\Zbb$ event purity is enhanced
by requiring that a $b$ tag exists in either event hemisphere.

  The lepton+tracks analysis proceeds by selecting identified leptons
($e$ or $\mu$) with large transverse momentum ($> 0.8$ GeV/c)
with respect to the nearest jet axis as a means to produce a sample enriched in
direct $(b \to l)$ decays.
A $B$ decay vertex is then reconstructed by intersecting all well-measured
tracks inside the jet with the lepton trajectory and by forming a weighted mean
intersection point. The weights are designed such as to enhance the
contribution from secondary tracks vs. primary tracks.
This method makes efficient use of the whole lepton sample, but has
slightly worse decay length and boost resolution than the lepton+``D'' method.

  The Charge Dipole analysis aims at reconstructing the $B$ and $D$ vertex
topologies in inclusive decays and tags the $B^0$ or $\overline{B^0}$
decay flavor based on the charge difference between the $B$ and $D$ vertices.
This analysis technique is unique to SLD.
Topological vertices with $M > 2$ GeV are selected as in the $B$ lifetime
analysis and the total track charge $Q$ is required to be 0
to enhance the fraction of $\bs$ decays in the sample and to increase the
quality of the charge difference reconstruction for neutral $B$ decays.
To select decays with non-negligible separation between the $B$ and $D$ decay
points, the probability for fitting all tracks to a single vertex is
required to be less than 5\%.
The tracks are then rearranged into various two-vertex combinations and
that with the lowest overall $\chi^2$ is selected. The vertex closer
to the IP is labelled ``$B$'' and that further away is labelled ``$D$.''
MC studies indicate that the track assignment to the $B$ ($D$) vertex
is 66\% (71\%) correct.
A ``Charge Dipole'' is defined as
$\delta Q \equiv D_{BD} \times SIGN (Q_D - Q_B)$,
where $D_{BD}$ is the distance between the two vertices and
$Q_B$ ($Q_D$) is the charge of the $B$ ($D$) vertex.
Positive (negative) values of $\delta Q$ tag $\overline{B^0}$ ($B^0$) decays
and the correct tag probability increases with increasing $|\delta Q|$
up to approximately 80\% for $\bs$ decays at least 1 mm away from the IP.

\begin{table}
\caption{\it\baselineskip=12pt
 Properties of the $\bsmix$ mixing analyses as extracted from the
 Monte Carlo simulation.
 The parameters $\sigma_{L1}$ and $\sigma_{L2}$ correspond to the
 Gaussian widths for fits to decay length residual distributions
 using a sum of two Gaussians with relative fractions $f_{L1}$ and
 $1-f_{L1}$.
 Similarly, the parameters $\sigma_{B1}$ and $\sigma_{B2}$ correspond to the
 Gaussian widths of the relative boost
 $\frac{\sigma_{\gamma\beta}}{\gamma\beta}$ residual distributions.
 }
\label{table_bsmix}
\vspace*{1mm}
\begin{center}
\begin{tabular}{llll}
   & lepton+``D'' & lepton+tracks & Charge Dipole \\
 \hline
 $\bs$ fraction         & ~~~~0.156    & ~~~~0.098 & ~~~~0.152 \\
 $udsc$ fraction        & ~~~~0.03     & ~~~~0.162 & ~~~~0.025 \\
 mistag rate            & ~~~~0.21     & ~~~~0.27  & ~~~~0.38  \\
 $\sigma_{L1}$ ($\mu$m) & ~~~~107      & ~~~~135   & ~~~~131   \\
 $f_{L1}$               & ~~~~0.60     & ~~~~0.70  & ~~~~0.60  \\
 $\sigma_{L2}$ ($\mu$m) & ~~~~330      & ~~~~614   & ~~~~500   \\
 $\sigma_{B1}$          & ~~~~0.070    & ~~~~0.073 & ~~~~0.080 \\
 $f_{B1}$               & ~~~~0.60     & ~~~~0.50  & ~~~~0.60  \\
 $\sigma_{B2}$          & ~~~~0.22     & ~~~~0.26  & ~~~~0.26  \\
 \hline
 \baselineskip=18pt
\end{tabular}
\end{center}
  \vspace*{-5mm}
\end{table}

  From the 1996-97 data, the number of events selected
is 1009, 4035, and 4634, in the lepton+``D'', lepton+tracks,
and Charge Dipole analyses, respectively.
The relevant properties of these samples have been extracted from the
simulation and are presented in Table~\ref{table_bsmix}.
The study of the time dependence of the oscillations is performed
with the amplitude method~\cite{Moser} in which the oscillation amplitude
$A$ is measured at a series of fixed $\Delta m_s$ values.
The oscillation probability becomes
$P_{B^0_s \rightarrow \overline{B^0_s}}(t) = \Gamma e^{-\Gamma t}
     \frac{1}{2} \left[ 1 - A \cos(\Delta m_s t) \right]$.
This method is equivalent to performing a Fourier transform analysis
and has the advantage of straightforward combination of several measurements
with correlated statistical and systematic uncertainties.

\begin{figure}[t]
  \hspace*{30mm}
  \epsfxsize9.6cm
  \epsfbox{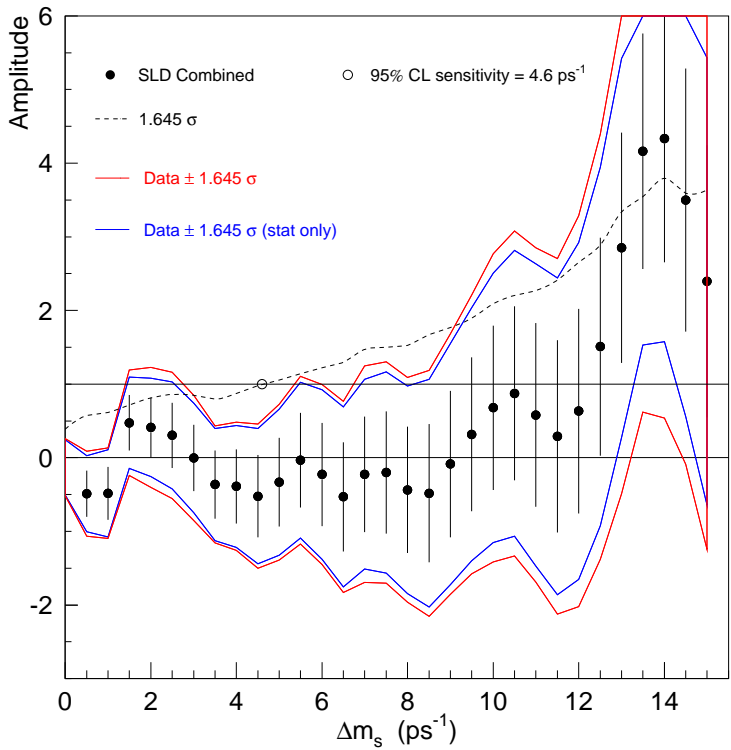}
  \caption{\it \label{fig_afit}
  \baselineskip=12pt
  Measured oscillation amplitude as a function of $\Delta m_s$ for
  all SLD analyses combined (points). Also shown is the $1.645\,\sigma$
  band around the measured amplitudes.}
  \baselineskip=18pt
\end{figure}

  The measured amplitudes for the three analyses are combined and
shown in Fig.~\ref{fig_afit}, taking
care of the statistical overlap between the two semileptonic methods
by removing 445 lepton+``D'' events from the lepton+tracks sample.
The dominant sources of systematic uncertainty have been examined.
The fraction of $\bs$ produced in the simulated $\Zbb$ decays was
varied according to $f_{\bs} = 0.115 \pm 0.020$,
and the resolution for both decay length and boost reconstruction
was varied by $\pm 10\%$.
The combined amplitude measurements allow the following regions to be
excluded at the 95\% C.L.: $\Delta m_s < 1.3$ ps$^{-1}$ and
$2.7 < \Delta m_s < 5.3$ ps$^{-1}$.
These regions obey the requirement $A + 1.645\,\sigma_A < 1$.
This is to be compared with an expected 95\% C.L. lower limit sensitivity
of 4.6 ps$^{-1}$ corresponding to the value of $\Delta m_s$ for
which $1.645\,\sigma_A = 1$.
The excellent proper time resolution results in a relatively small
growth of $\sigma_A$ with increasing $\Delta m_s$, indicating
that SLD will be sensitive to high values of $\Delta m_s$ with a
modest increase in statistics.
Studies show that a 95\% C.L. lower limit sensitivity of 20 ps$^{-1}$
can be achieved with a million hadronic $Z^0$ decays.

\vspace*{3mm}
\noindent {\large \bf  4.  Search for $\bsg$ Decays}
\vspace*{3mm}

  It has been suggested that the long-standing puzzle of the low
semileptonic branching ratio in conjunction with a low charm yield
could be resolved by an enhanced branching ratio ($\sim 10\%$ rather than
0.2\% in the SM) for transitions of the type $\bsg$~\cite{Kagan}.
Such a large branching ratio would be clearly visible in the $B$ decay
kaon spectrum at high momentum~\cite{Kagan}.
SLD has searched for an enhancement in the $B$ decay kaon momentum
transverse to the $B$ flight direction using
the 1993-95 data sample of 150,000 hadronic $Z^0$ decays~\cite{Daoudi}.
The search concentrates on a sample of topological vertices containing
kaons identified in the Cherenkov Ring Imaging
Detector and with transverse momentum $p_T > 1.8$ GeV/c.
Vertices are required to have $M > 2$ GeV and vertex fit
probability greater than 5\% to enhance the fraction of signal events.
The difference between the number of kaons in the data and the MC
(which includes only the dominant $b \to c$ transition) was found to be
$12.9 \pm 5.9$(stat) $ \pm 3.1$(syst).
This analysis has been updated with another 150,000 hadronic $Z^0$ decays
collected in 1996-97 and the (data-MC) difference is found to be
$3.6 \pm 4.5$(stat) $ \pm 2.5$(syst) in that sample.
As a check of the detector simulation,
the $p_T$ distributions of identified muons attached to the vertex
for data and MC are compared and found to agree well.
Furthermore, good overall agreement between data and MC is obtained in
the kaon $p_T$ distribution for vertices with fit
probabilities $< 5\%$---a sample significantly enhanced in $B$ decays
to single- and double-charm final states.

\begin{center}
\begin{figure}[t]
  \hspace*{20mm}
  \epsfxsize12cm
  \epsfbox{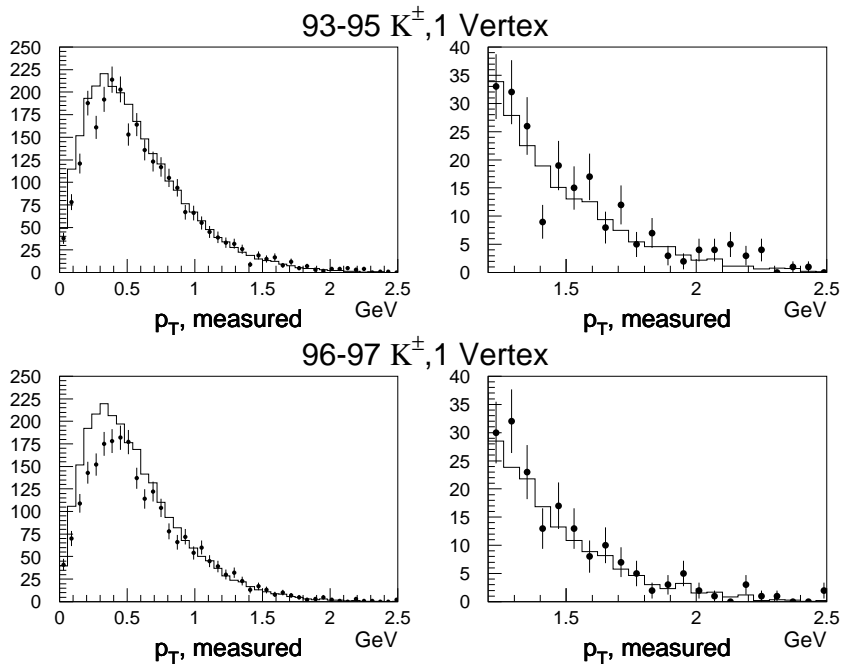}
  \caption{\it \label{fig_kaonpt}
  \baselineskip=12pt
  Distributions of kaon transverse momentum for
  1993-95 data (points) and Monte Carlo (histograms) at top,
  and for 1996-97 data (points) and Monte Carlo (histograms) at bottom.
  The distributions on the right show the high-$p_T$ region only.}
  \baselineskip=18pt
\end{figure}
\end{center}

Thus, no significant enhancement is observed in the data
(see Fig.~\ref{fig_kaonpt})
indicating no evidence for enhanced $\bsg$ transitions.

%

\end{document}